\documentclass[]{article}
\usepackage{lmodern}
\usepackage{amssymb,amsmath}
\usepackage{ifxetex,ifluatex}
\usepackage{fixltx2e} 
\ifnum 0\ifxetex 1\fi\ifluatex 1\fi=0 
  \usepackage[T1]{fontenc}
  \usepackage[utf8]{inputenc}
\else 
  \ifxetex
    \usepackage{mathspec}
  \else
    \usepackage{fontspec}
  \fi
  \defaultfontfeatures{Ligatures=TeX,Scale=MatchLowercase}
\fi
\IfFileExists{upquote.sty}{\usepackage{upquote}}{}
\IfFileExists{microtype.sty}{%
\usepackage{microtype}
\UseMicrotypeSet[protrusion]{basicmath} 
}{}
\usepackage[margin=1in]{geometry}
\usepackage{hyperref}
\hypersetup{unicode=true,
            pdftitle={Facilitating Ontology Development with Continuous Evaluation},
            pdfauthor={Dejan Lavbič and Marjan Krisper},
            pdfborder={0 0 0},
            breaklinks=true}
\usepackage{natbib}
\usepackage{longtable,booktabs}
\usepackage{graphicx,grffile}
\makeatletter
\def\maxwidth{\ifdim\Gin@nat@width>\linewidth\linewidth\else\Gin@nat@width\fi}
\def\maxheight{\ifdim\Gin@nat@height>\textheight\textheight\else\Gin@nat@height\fi}
\makeatother
\setkeys{Gin}{width=\maxwidth,height=\maxheight,keepaspectratio}
\IfFileExists{parskip.sty}{%
\usepackage{parskip}
}{
\setlength{\parindent}{0pt}
\setlength{\parskip}{6pt plus 2pt minus 1pt}
}
\providecommand{\tightlist}{%
  \setlength{\itemsep}{0pt}\setlength{\parskip}{0pt}}
\ifx\paragraph\undefined\else
\let\oldparagraph\paragraph
\renewcommand{\paragraph}[1]{\oldparagraph{#1}\mbox{}}
\fi
\ifx\subparagraph\undefined\else
\let\oldsubparagraph\subparagraph
\renewcommand{\subparagraph}[1]{\oldsubparagraph{#1}\mbox{}}
\fi

\let\rmarkdownfootnote\footnote%
\def\footnote{\protect\rmarkdownfootnote}

\usepackage{titling}

  \title{Facilitating Ontology Development with Continuous Evaluation}
    \pretitle{\vspace{\droptitle}\centering\huge}
  \posttitle{\par}
    \author{Dejan Lavbič and Marjan Krisper}
    \preauthor{\centering\large\emph}
  \postauthor{\par}
    \date{}
    \predate{}\postdate{}

\usepackage{booktabs}
\usepackage{longtable}
\usepackage{array}
\usepackage{multirow}
\usepackage[table]{xcolor}
\usepackage{wrapfig}
\usepackage{float}
\usepackage{colortbl}
\usepackage{pdflscape}
\usepackage{tabu}
\usepackage{threeparttable}
\usepackage{threeparttablex}
\usepackage[normalem]{ulem}
\usepackage{makecell}

\usepackage{amsthm}

\theoremstyle{definition}

\theoremstyle{definition}

\theoremstyle{definition}

\theoremstyle{remark}

\let\BeginKnitrBlock\begin \let\EndKnitrBlock\end
\begin{document}
\maketitle

\begin{quote}
\textbf{Dejan Lavbič} and Marjan Krisper. 2010. \textbf{Facilitating
Ontology Development with Continuous Evaluation},
\href{https://www.mii.lt/informatica/}{Informatica \textbf{(INFOR)}},
21(4), pp.~533 - 552.
\end{quote}

\section*{Abstract}\label{abstract}
\addcontentsline{toc}{section}{Abstract}

In this paper we propose facilitating ontology development by constant
evaluation of steps in the process of ontology development. Existing
methodologies for ontology development are complex and they require
technical knowledge that business users and developers don't poses. By
introducing ontology completeness indicator developer is guided
throughout the development process and constantly aided by
recommendations to progress to next step and improve the quality of
ontology. In evaluating the ontology, several aspects are considered;
from description, partition, consistency, redundancy and to anomaly. The
applicability of the approach was demonstrated on Financial Instruments
and Trading Strategies (FITS) ontology with comparison to other
approaches.

\section*{Keywords}\label{keywords}
\addcontentsline{toc}{section}{Keywords}

Ontology development methodology, ontology evaluation, ontology
completeness, rapid ontology development, semantic web

\section{Introduction}\label{introduction}

The adoption of Semantic Web technologies is less than expected and is
mainly limited to academic environment. We are still waiting for wide
adoption in industry. We could seek reasons for this in technologies
itself and also in the process of development, because existence of
verified approaches is a good indicator of maturity. As technologies are
concerned there are numerous available for all aspects of Semantic Web
applications; from languages for capturing the knowledge, persisting
data, inferring new knowledge to querying for knowledge etc. In the
methodological sense there is also a great variety of methodologies for
ontology development available, as it will be further discussed in
section \ref{related-work}, but the simplicity of using approaches for
ontology construction is another issue. Current approaches in ontology
development are technically very demanding and require long learning
curve and are therefore inappropriate for developers with little
technical skills and knowledge. In majority of existing approaches an
additional role of knowledge engineer is required for mediation between
actual knowledge that developers possess and ontology engineers who
encode knowledge in one of the selected formalisms. The use of business
rules management approach \citep{smaizys_business_2009} seems like an
appropriate way to simplification of development and use of ontologies
in business applications. Besides simplifying the process of ontology
creation we also have to focus on very important aspect of ontology
completeness. The problem of error-free ontologies has been discussed in
\citep{fahad_ontological_2008, porzel_task-based_2004} and several types
of errors were identified - inconsistency, incompleteness, redundancy,
design anomalies etc. All of these problems have to already be addressed
in the development process and not only after development has reached
its final steps.

In this paper we propose a Rapid Ontology Development (ROD) approach
where ontology evaluation is performed during the whole lifecycle of the
development. The idea is to enable developers to rather focus on the
content than the formalisms for encoding knowledge. Developer can
therefore, based on recommendations, improve the ontology and eliminate
the error or bad design. It is also a very important aspect that, before
the application, the ontology is error free. Thus we define ROD model
that introduces detail steps in ontology manipulation. The starting
point was to improve existing approaches in a way of simplifying the
process and give developer support throughout the lifecycle with
continuous evaluation and not to conclude with developed ontology but
enable the use of ontology in various scenarios. By doing that we try to
achieve two things:

\begin{itemize}
\tightlist
\item
  guide developer through the process of ontology construction and
\item
  improve the quality of developed ontology.
\end{itemize}

The remainder of the paper is structured as follows. In the following
section \ref{related-work} state of the art is presented with the review
of existing methodologies for ontology development and approaches for
ontology evaluation. After highlighting some drawbacks of current
approaches section \ref{ROD} presents the ROD approach. Short overview
of the process and stages is given with the emphasis on ontology
completeness indicator. The details of ontology evaluation and ontology
completeness indicator are given in section \ref{indicator}, where all
components (description, partition, redundancy and anomaly) that are
evaluated are presented. In section \ref{evaluation} evaluation and
discussion about the proposed approach according to the results obtained
in the experiment of \textbf{Financial Instruments and Trading
Strategies (FITS)} is presented. Finally in section
\ref{conclusion-and-future-work} conclusions with future work are given.

\section{Related work}\label{related-work}

\subsection{Review of related
approaches}\label{review-of-related-approaches}

Ontology is a vocabulary that is used for describing and presentation of
a domain and also the meaning of that vocabulary. The definition of
ontology can be highlighted from several aspects. From taxonomy
\citep{corcho_methodologies_2003, sanjuan_text_2006, veale_analogy-oriented_2006}
as knowledge with minimal hierarchical structure, vocabulary
\citep{bechhofer_thesaurus_2001, miller_wordnet:_1995} with words and
synonyms, topic maps \citep{dong_hyo-xtm:_2004, park_xml_2002} with the
support of traversing through large amount of data, conceptual model
\citep{jovanovic_achieving_2005, mylopoulos_information_1998} that
emphasizes more complex knowledge and logic theory
\citep{corcho_methodologies_2003, dzemyda_optimization_2009, waterson_verifying_1999}
with very complex and consistent knowledge.

Ontologies are used for various purposes such as natural language
processing \citep{staab_system_1999}, knowledge management
\citep{davies_semantic_2006}, information extraction
\citep{wiederhold_mediators_1992}, intelligent search engines
\citep{heflin_searching_2000}, digital libraries
\citep{kesseler_schema_1996}, business process modeling
\citep{brambilla_software_2006, ciuksys_reusing_2007, magdalenic_dynamic_2009}
etc. While the use of ontologies was primarily in the domain of
academia, situation now improves with the advent of several
methodologies for ontology manipulation. Existing methodologies for
ontology development in general try to define the activities for
ontology management, activities for ontology development and support
activities. Several methodologies exist for ontology manipulation and
will be briefly presented in the following section. CommonKADS
\citep{schreiber_knowledge_1999} is in fact not a methodology for
ontology development, but is focused towards knowledge management in
information systems with analysis, design and implementation of
knowledge. CommonKADS puts an emphasis to early stages of software
development for knowledge management. Enterprise Ontology
\citep{uschold_towards_1995} recommends three simple steps: definition
of intention; capturing concepts, mutual relation and expressions based
on concepts and relations; persisting ontology in one of the languages.
This methodology is the groundwork for many other approaches and is also
used in several ontology editors. METHONTOLOGY
\citep{fernandez-lopez_building_1999} is a methodology for ontology
creation from scratch or by reusing existing ontologies. The framework
enables building ontology at conceptual level and this approach is very
close to prototyping. Another approach is TOVE
\citep{uschold_ontologies:_1996} where authors suggest using
questionnaires that describe questions to which ontology should give
answers. That can be very useful in environments where domain experts
have very little expertise of knowledge modeling. Moreover authors of
HCONE \citep{kotis_human_2003} present decentralized approach to
ontology development by introducing regions where ontology is saved
during its lifecycle. OTK Methodology \citep{sure_methodology_2003}
defines steps in ontology development into detail and introduces two
processes -- Knowledge Meta Process and Knowledge Process. The steps are
also supported by a tool. UPON \citep{nicola_building_2005} is an
interesting methodology that is based on Unified Software Development
Process and is supported by UML language, but it has not been yet fully
tested. The latest proposal is DILIGENT \citep{davies_semantic_2006} and
is focused on different approaches to distributed ontology development.

From information systems development point of view there are several
methodologies that share similar ideas found in ontology development.
Rapid Ontology Development model, presented in this paper follows
examples mainly from blended, object-oriented, rapid development and
people-oriented methodologies \citep{avison_information_2006}. In
blended methodologies, that are formed from (the best) parts of other
methodologies, the most influential for our approach was Information
Engineering \citep{martin_information_1981} that is viewed as a
framework within which a variety of techniques are used to develop good
quality information systems in an efficient way. In object-oriented
approaches there are two representatives -- Object-Oriented Analysis
(OOA; \citet{booch_object_1993}) and Rational Unified Process (RUP;
\citet{jacobson_unified_1999}). Especially OOA with its five major
activities: finding class and objects, identifying structures,
indentifying subjects, defining attributes and defining services had
profound effect on our research, while it was extended with the support
of design and implementation phases that are not included in OOA. The
idea of rapid development methodologies is closely related to ROD
approach and current approach addresses the issue of rapid ontology
development which is based on rapid development methodologies of
information systems. James Martin's RAD \citep{martin_rapid_1991} is
based on well known techniques and tools but adopts prototyping approach
and focuses on obtaining commitment from the business users. Another
rapid approach is Dynamic Systems Development Method (DSDM;
\citet{consortium_dsdm_2005}) which has some similarities with Extreme
Programming (XP; \citet{beck_extreme_2004}). XP attempts to support
quicker development of software, particularly for small and medium-sized
applications.

Comparing to techniques involved in information systems development, the
ontology development in ROD approach is mainly based on \emph{holistic
techniques} (rich pictures, conceptual models, cognitive mapping),
\emph{data techniques} (entity modeling, normalization), \emph{process
techniques} (decision trees, decision tables, structured English) and
\emph{project management techniques} (estimation techniques).

The ROD approach extends reviewed methodologies by simplifying
development steps and introducing continuous evaluation of developed
ontology. This is achieved by ontology completeness indicator that is
based on approaches for ontology evaluation. Based on existing reviews
in
\citep{brank_survey_2005, gangemi_modelling_2006, gomez-perez_evaluation_1999, hartmann_d1.2.3_2004}
we classify evaluation approaches into following categories:

\begin{itemize}
\tightlist
\item
  compare ontology to \emph{``golden standard''}
  \citep{maedche_measuring_2002},
\item
  using ontology in an \emph{application} and evaluating results
  \citep{porzel_task-based_2004},
\item
  compare with source of data about the \emph{domain to be covered} by
  ontology \citep{brewster_data_2004} and
\item
  \emph{evaluation} done \emph{by humans}
  \citep{lozano-tello_ontometric:_2004, noy_user_2005}.
\end{itemize}

Usually evaluation of different levels of ontology separately is more
practical than trying to directly evaluate the ontology as whole.
Therefore, classification of evaluation approaches based on the level of
evaluation is also feasible and is as follows: lexical, vocabulary or
data layer, hierarchy or taxonomy, other semantic relations, context or
application level, syntactic level, structure, architecture and design.
Prior the application of ontologies we have to assure that they are free
of errors. The research performed by \citet{fahad_ontological_2008}
resulted in classification and consequences of ontology errors. These
errors can be divided into inconsistency errors, incompleteness errors,
redundancy errors and design anomalies.

\subsection{Problem and proposal for
solution}\label{problem-and-proposal-for-solution}

The review of existing approaches for ontology development in this
section pointed out that several drawbacks exist. Vast majority of
ontology development methodologies define a complex process that demands
a long learning curve. The required technical knowledge is very high
therefore making ontology development very difficult for nontechnically
oriented developers. Among methodologies for ontology development there
is a lack of rapid approaches which can be found in traditional software
development approaches. On the other hand methodologies for traditional
software development also fail to provide sufficient support in ontology
development. This fact can be confirmed with the advent of several
ontology development methodologies presented at the beginning of this
section. Majority of reviewed methodologies also include very limited
evaluation support of developed ontologies. If this support exists it is
limited to latter stages of development and not included throughout the
process.

This paper introduces a novel approach in ontology modeling based on
good practices and existing approaches
\citep{allemang_semantic_2008, cardoso_semantic_2007, fahad_ontological_2008, fernandez-lopez_building_1999, sure_methodology_2003, uschold_towards_1995}
while trying to minimize the need of knowing formal syntax required for
codifying the ontology and therefore bringing ontology modeling closer
to business users who are actual knowledge holders. Based on the
findings from the comparison of existing methodologies for ontology
development and several evaluation approaches it has been noted that no
approach exist that would constantly evaluate ontology during its
lifecycle. The idea of proposed ROD approach with ontology completeness
evaluation presented in section \ref{ROD} is to create a feedback loop
between developed ontology and its completeness by introducing indicator
for completeness. With ROD approach detailed knowledge of development
methodology is also not required as the process guides developers
through the steps defined in methodology. By extending existing
approaches with constant evaluation the quality of final artifact is
improved and the time for development is minimized as discussed in
section \ref{indicator}.

\section{Rapid Ontology Development}\label{ROD}

\subsection{Introduction to ROD
process}\label{introduction-to-rod-process}

The process for ontology development ROD (Rapid Ontology Development)
that we propose is based on existing approaches and methodologies (see
section \ref{related-work}) but is enhanced with continuous ontology
evaluation throughout the complete process. It is targeted at domain
users that are not familiar with technical background of constructing
ontologies.

Developers start with capturing concepts, mutual relations and
expressions based on concepts and relations. This task can include
reusing elements from various resources or defining them from scratch.
When the model is defined, schematic part of ontology has to be binded
to existing instances of that vocabulary. This includes data from
relational databases, text file, other ontologies etc. The last step in
bringing ontology into use is creating functional components for
employment in other systems.

\subsection{ROD stages}\label{ROD-stages}

\begin{figure}

{\centering \includegraphics[width=0.7\linewidth]{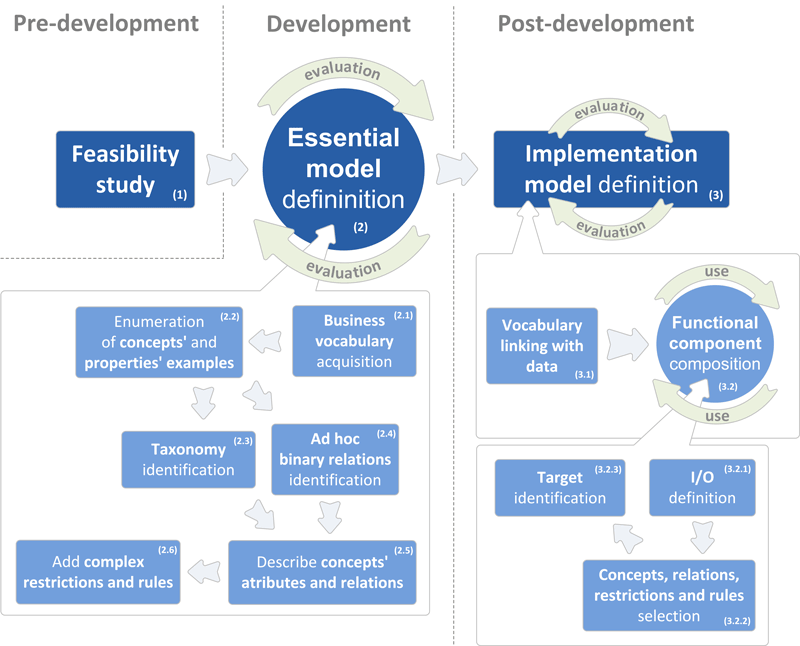}

}

\caption{Process of rapid ontology development (ROD)}\label{fig:ROD-process}
\end{figure}

The ROD development process can be divided into the following stages:
\emph{pre-development}, \emph{development} and \emph{post-development}
as depicted in Figure \ref{fig:ROD-process}. Every stage delivers a
specific output with the common goal of creating functional component
based on ontology that can be used in several systems and scenarios. In
pre-development stage the output is feasibility study that is used in
subsequent stage development to construct essential model definition.
The latter artifact represents the schema of problem domain that has to
be coupled with instances from the real world. This is conducted in the
last stage post-development which produces functional component for
usage in various systems.

\begin{figure}

{\centering \includegraphics[width=0.3\linewidth]{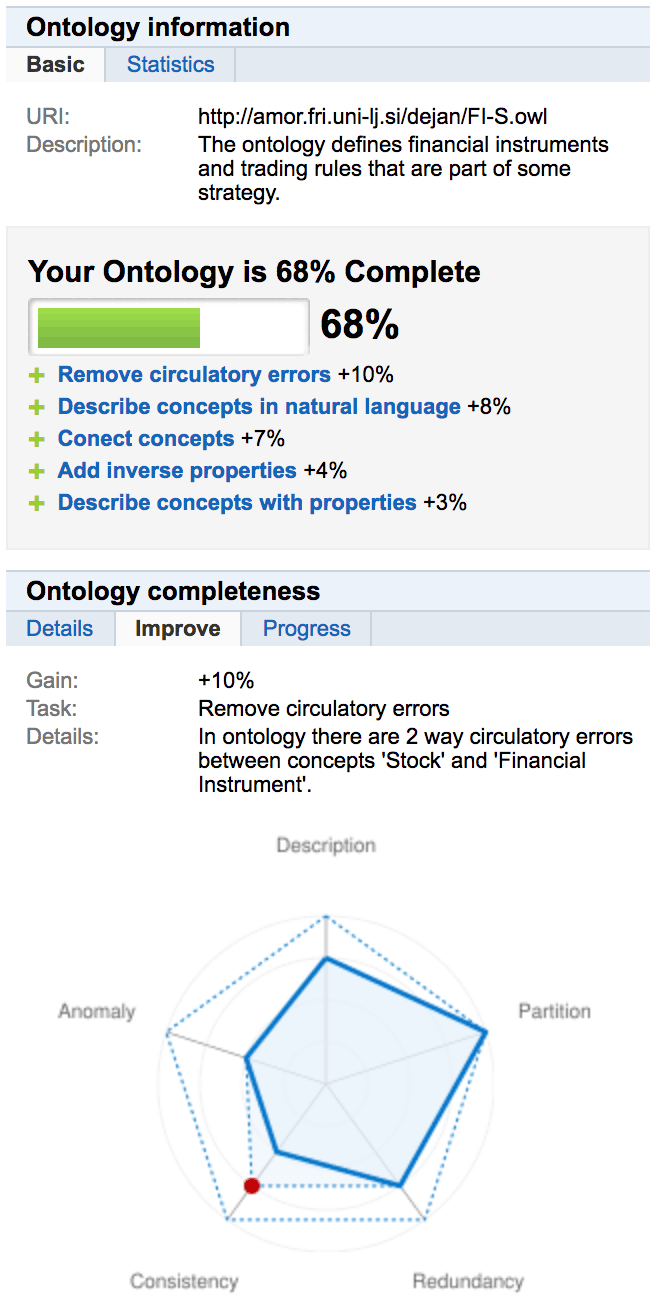}

}

\caption{Display of ontology completeness (OC) results and improvement recommendations}\label{fig:OC-GUI}
\end{figure}

The role of constant evaluation as depicted in Figure
\ref{fig:ROD-process} is to guide developer in progressing through steps
of ROD process or it can be used independently of ROD process. In latter
case, based on semantic review of ontology, enhancements for ontology
improvement are available to the developer in a form of multiple actions
of improvement, sorted by their impact. Besides actions and their
impacts, detail explanation of action is also available (see Figure
\ref{fig:OC-GUI}).

In case of following ROD approach, while developer is in a certain step
of the process, the OC measurement is adapted to that step by
redefinition of weights (see Figure \ref{fig:OC-weights} for
distribution of weights by ROD steps) for calculation (e.g., in Step 2.1
of ROD process where business vocabulary acquisition is performed, there
is no need for semantic checks like instance redundancy, lazy concept
existence or inverse property existence, but the emphasis is rather on
description of TBox and RBox component and path existence between
concepts).

\begin{figure}

{\centering \includegraphics[width=0.7\linewidth]{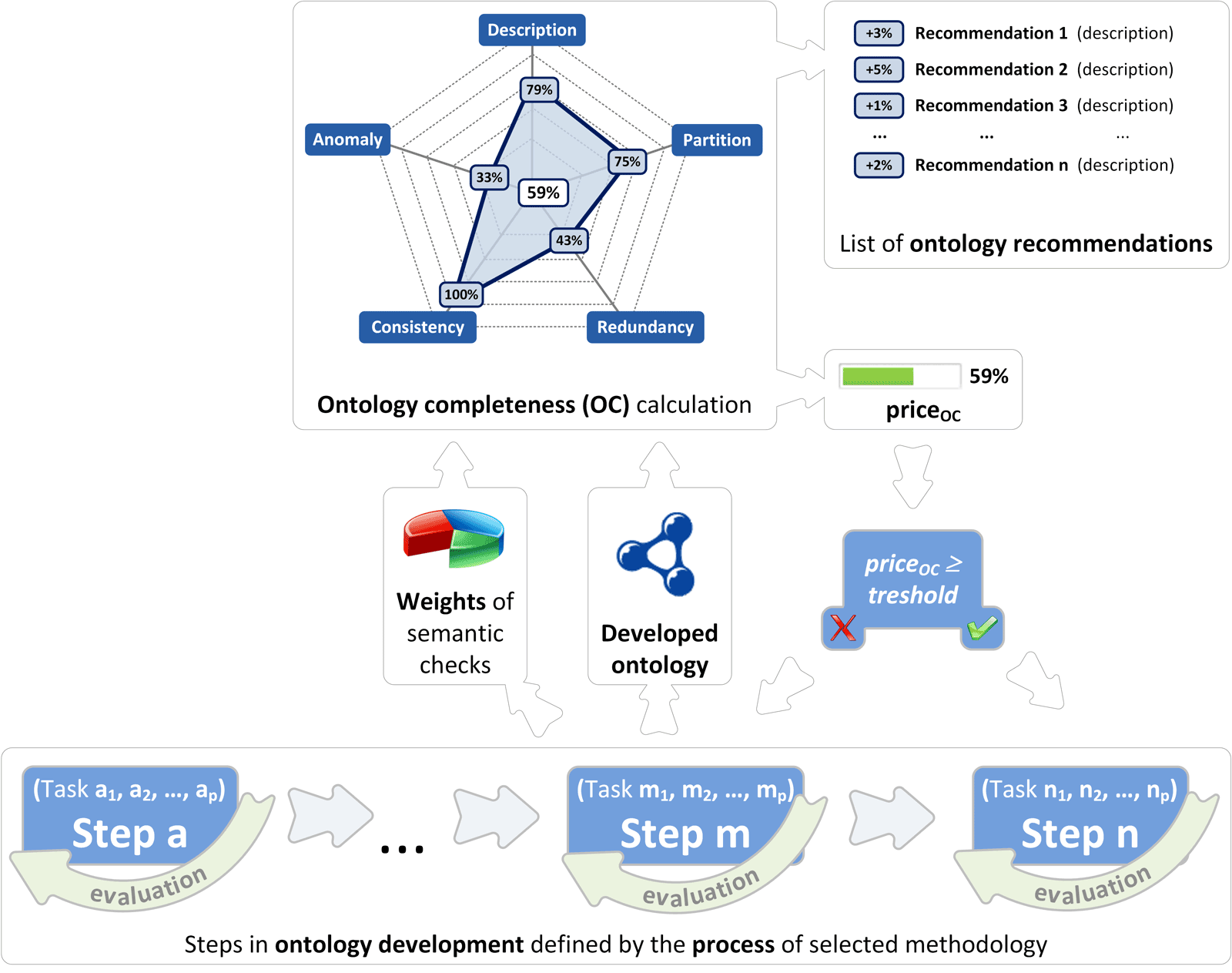}

}

\caption{OC calculation}\label{fig:OC-calculation}
\end{figure}

When OC measurement reaches a threshold (e.g., \(80\%\)) developer can
progress to the following step (see Figure \ref{fig:OC-calculation}).
The adapted OC value for every phase is calculated on-the-fly and
whenever a threshold value is crossed, a recommendation for progressing
to next step is generated. This way developer is aided in progressing
through steps of ROD process from business vocabulary acquisition to
functional component composition.

In case that ontology already exists, with OC measure we can place the
completeness of ontology in ROD process and start improving ontology in
suggested phase of development (e.g., ontology has taxonomy already
defined, so we can continue with step 2.4 where ad hoc binary relations
identification takes place).

\subsection{Ontology evaluation and ontology completeness
indicator}\label{indicator}

\textbf{Ontology completeness (OC)} indicator used for guiding developer
in progressing through steps of ROD process and ensuring the required
quality level of developed ontology is defined as

\begin{equation}
OC = f \left( C, P, R, I \right) \in [0, 1]
\label{eq:OC}
\end{equation}

where \(C\) is set of concepts, \(P\) set of properties, \(R\) set of
rules and \(I\) set of instances. Based on these input the output value
in an interval \([0, 1]\) is calculated. The higher the value, more
complete the ontology is. OC is weighted sum of semantic checks, while
weights are being dynamically altered when traversing from one phase in
ROD process to another. OC can be further defined as

\begin{equation}
OC = \sum_{i=1}^{n} w_i^{'} \cdot leafCondition_i
\label{eq:OC-sum}
\end{equation}

where \(n\) is the number of leaf conditions and \(leafCondition\) is
leaf condition, where semantic check is executed. For relative weights
and leaf condition calculation the following restrictions apply
\(\sum_i w_i^{'} = 1\), \(\forall w_i^{'} \in [0, 1]\) and
\(\forall leafCondition_i \in [0, 1]\). Relative weight \(w_i^{'}\)
denotes global importance of \(leafCondition_i\) and is dependent on all
weights from leaf to root concept.

\begin{figure}

{\centering \includegraphics[width=0.7\linewidth]{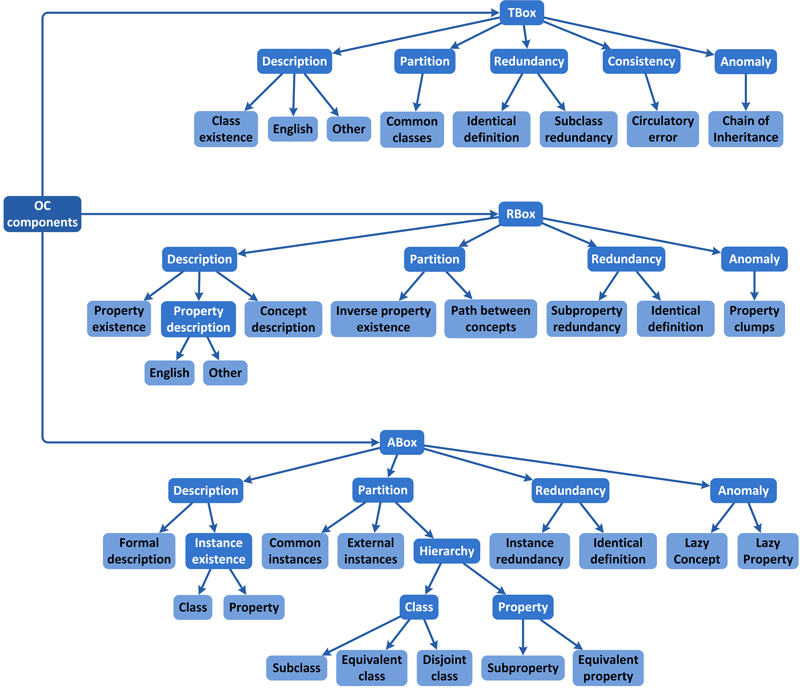}

}

\caption{Ontology completeness (OC) tree of conditions, semantic checks and corresponding weights}\label{fig:OC-tree}
\end{figure}

The tree of conditions in OC calculation is depicted in Figure
\ref{fig:OC-tree} and contains semantic checks that are executed against
the ontology. The top level is divided into \emph{TBox}, \emph{RBox} and
\emph{ABox} components. Subsequent levels are then furthermore divided
based on ontology error classification \citep{fahad_ontological_2008}.
Aforementioned sublevels are \emph{description}, \emph{partition},
\emph{redundancy}, \emph{consistency} and \emph{anomaly}.

This proposed structure can be easily adapted and altered for custom
use. Leafs in the tree of OC calculation are implemented as semantic
checks while all preceding elements are aggregation with appropriate
weights. Algorithm for ontology completeness (OC) price is depicted in
Definition \ref{def:OC-evaluation}, where \(X\) is condition and
\(w = w(X, Y)\) is the weight between condition \(X\) and condition
\(Y\).

\BeginKnitrBlock{definition}[Ontology completeness evaluation algorithm]
\protect\hypertarget{def:OC-evaluation}{}{\label{def:OC-evaluation}
\iffalse (Ontology completeness evaluation algorithm) \fi{} }

\begin{align*}
&\text{' Evaluation is executed on top condition "OC components" with weight 1} \\
&\textbf{Evaluate } \boldsymbol{(X, w)} \\
&\quad price_{OC} = 0 \\
&\quad \text{mark condition } X \text{ as visited} \\
&\quad \text{if not exists sub-condition of } X \\
&\qquad \text{' Execute semantic check on leaf element} \\
&\qquad \text{return } w \cdot exec(X) \\
&\quad \text{else for all conditions } Y \text{ that are sub-conditions of } X \text{ such that } Y \text{ is not visited} \\
&\qquad \text{' Aggregate ontology evaluation prices} \\
&\qquad \text{if } w(X, Y) \neq 0 \\
&\qquad \quad price_{OC} = price_{OC} + Evaluate(Y, w(X, Y)) \\
&\quad \text{return } w \cdot price_{OC} \\
&\textbf{End}
\end{align*}
\EndKnitrBlock{definition}

Each leaf condition implements a semantic check against ontology and
returns value \(leafCondition \in [0, 1]\).

\begin{figure}

{\centering \includegraphics[width=0.7\linewidth]{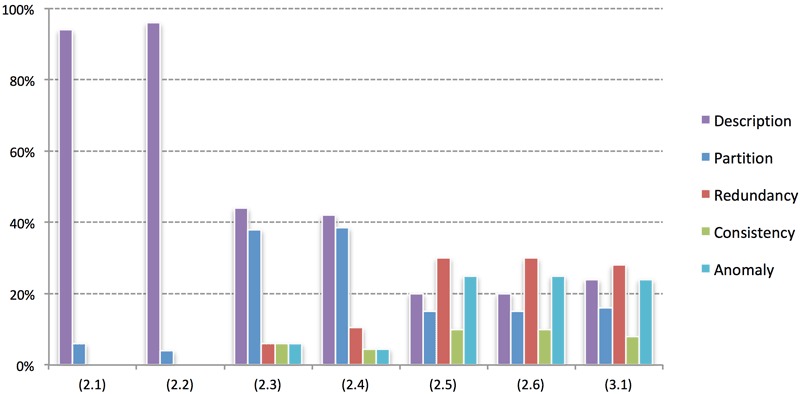}

}

\caption{Impact of weights on OC sublevels in ROD process}\label{fig:OC-weights}
\end{figure}

Figure \ref{fig:OC-weights} depicts the distribution of OC components
(description, partition, redundancy, consistency and anomaly) regarding
individual phase in ROD process (see section \ref{ROD-stages}). In first
two phases 2.1 and 2.2 developer deals with business vocabulary
identification and enumeration of concepts' and properties' examples.
Evidently with aforementioned steps emphasis is on description of
ontology, while partition is also taken into consideration. The
importance of components description and partition is then in latter
steps decreased but it still remains above average. In step 2.3 all
other components are introduced (redundancy, consistency and anomaly),
because developer is requested to define taxonomy of schematic part of
ontology. While progressing to the latter steps of ROD process emphasis
is on detail description of classes, properties and complex restriction
and rules are also added. At this stage redundancy becomes more
important. This trend of distributions of weights remains similarly
balanced throughout the last steps 2.5 and 2.6 of development phase. In
post-development phase when functional component composition is
performed, ontology completeness calculation is mainly involved in
redundancy, description and anomaly checking. The details about
individual OC components are emphasized and presented in details in the
following subsections.

\subsubsection{Description}\label{description}

Description of ontology's components is very important aspect mainly in
early stages of ontology development. As OC calculation is concerned
there are several components considered:

\begin{itemize}
\tightlist
\item
  \emph{existence of entities} (classes and properties) and
  \emph{instances},
\item
  (multiple) \emph{natural language descriptions} of TBox and RBox
  components and
\item
  \emph{formal description} of concepts and instances.
\end{itemize}

The notion of existence of entities is very straightforward; if ontology
doesn't contain any entities than we have no artifacts to work with.
Therefore the developer is by this metric encouraged to first define
schematic part of ontology with classes and properties and then also to
add elements of ABox component in a form of individuals.

Next aspect is natural language descriptions of entities. This element
is despite of its simplicity one of the most important, due to ability
to include these descriptions in further definition of complex axioms
and rules \citep{vasilecas_towards_2009}. Following business rules
approach \citep{vasilecas_practical_2008} it's feasible to create
templates for entering this data on-the-fly by employing this natural
description of entities. Developer is encouraged to describe all
entities (classes and properties) with natural language using readable
labels (e.g., \texttt{rdfs:label} and \texttt{rdfs:comment}) that don't
add to the meaning of captured problem domain but greatly improves human
readability of defined ontology. When constructing ontology it is always
required to provide labels and description in English, but the use of
other languages is also recommended to improve employment of ontology.

The last aspect of ontology description is formal description of TBox
and ABox components that concerns concepts and instances. When
describing classes with properties ontologists tend to forget defining
domain and range values. This is evaluated for schematic part of
ontology while for instances all required axioms are considered that are
defined in TBox or ABox. Ontologists tend to leave out details of
instances that are required (e.g., cardinality etc.).

\subsubsection{Partition}\label{partition}

Partition errors deal with omitting important axioms or information
about the classification of concept and therefore reducing reasoning
power and inferring mechanisms. In OC calculation several components are
considered:

\begin{itemize}
\tightlist
\item
  \emph{common classes} and \emph{instances},
\item
  \emph{external instances} of ABox component,
\item
  \emph{connectivity of concepts} of TBox component and
\item
  \emph{hierarchy of entities}.
\end{itemize}

The notion of common classes deals with the problem of defining a class
that is a sub-class of classes that are disjoint. The solution is to
check every class \(C_i\) if exist super-classes \(C_j\) and \(C_k\)
that are disjoint. Similar is with common instances where situation can
occur where instance is member of disjointing classes.

When decomposing classes in sub-class hierarchy it is often the case
that super-class instance is not a member of any sub-class. In that case
we deal with a problem of external instances. The solution is to check
every class \(C_i\) if exist any instance that is a member of \(C_i\),
but not a member of any class in set of sub-classes.

The aspect of connectivity of concepts deals with ontology as whole and
therefore not allowing isolated parts that are mutually disconnected.
The first semantic check deals with existence of inverse properties. If
we want to contribute to full traversal among classes in ontology the
fact that every object property has inverse property defined is very
important.

The second semantic check deals with existence of path between concepts.
Ontology is presented as undirected graph \(G = (V, E)\) and we try to
identify maximum disconnected graphs.

The last aspect of ontology completeness as partition is concerned with
hierarchy of entities. We introduce data oriented approach for
definition of hierarchy of entities where technical knowledge from
domain user is not required. This is based on requirement that for every
class and property defined ontologist is requested to insert also few
instances (see preliminary steps in ROD process introduced in section
\ref{ROD-stages}). After this requirement is met, set of competency
questions are introduced to the domain user and the result are
automatically defined hierarchy axioms (e.g., \texttt{rdfs:subClassOf},
\texttt{owl:equivalentClass}, \texttt{owl:disjointWith},
\texttt{rdfs:subPropertyOf} and \texttt{rdfs:equivalentProperty}).

The approach for disjoint class recommendation is depicted in Definition
\ref{def:disjoint-axiom}, while approach for other hierarchy axioms is
analogous.

\BeginKnitrBlock{definition}[Recommend disjoint axiom between classes]
\protect\hypertarget{def:disjoint-axiom}{}{\label{def:disjoint-axiom}
\iffalse (Recommend disjoint axiom between classes) \fi{} }

\begin{align*}
&\textbf{recommendDisjointWithClasses} \\
&\quad \tau_{\subseteq}^{sibling} = \{ \} \leftarrow \text{ Set of all sub-class pairs } (C, D) \\
&\quad Q_n \leftarrow \text{ Competency questions} \\
&\quad disjointClassRecommend = \{ \} \\
&\quad \text{for each } C_i \in TBox \\
&\qquad \text{add all sub-class pairs of class } C_i \text{ to } \tau_{\subseteq}^{sibling} \\
&\qquad \text{for each sub-class pair } (C_j, C_k) \in TBox \text{ where } C_j \subseteq C_i \wedge C_k \subseteq C_i \wedge C_j \neq C_k \\
&\qquad \quad \text{if } \exists i (C_j), i (C_k) \in ABox : ( \neg Q_1 (C_j, C_k) \wedge \neg Q_3 (C_j, C_k) ) \text{ then} \\
&\qquad \qquad \text{if } C_j \cap C_k \neq \{ \} \text{ then} \\
&\qquad \qquad \quad disjointClassRecommend = disjointClassRecommend \cup (C_j, C_k) \\
&\qquad \qquad \text{end if} \\
&\qquad \quad \text{end if} \\
&\qquad \text{end for} \\
&\quad \text{end for} \\
&\quad price = 1 - \frac{ \left | disjointClassRecommend \right | } { \left | \tau_{\subseteq}^{sibling} \right | } \\
&\quad \text{return } \boldsymbol{disjointClassRecommend} \text{ and } \boldsymbol{price} \\
&\textbf{end}
\end{align*}
\EndKnitrBlock{definition}

Using this approach of recommendation, domain users can define axioms in
ontology without technical knowledge of ontology language, because with
data driven approach (using instances) and competency questions the OC
calculation indicator does that automatically.

Redundancy occurs when particular information is inferred more than once
from entities and instances. When calculating OC we take into
consideration following components:

\begin{itemize}
\tightlist
\item
  \emph{identical formal definition} and
\item
  \emph{redundancy in hierarchy of entities}.
\end{itemize}

When considering identical formal definition, all components (TBox, RBox
and ABox) have to be checked. For every entity or instance Ai all
belonging axioms are considered. If set of axioms of entity or instance
\(A_i\) is identical to set of axioms of entity or instance \(A_j\) and
\(A_i \neq A_j\), then entities or instances \(A_i\) and \(A_j\) have
identical formal definition. This signifies that \(A_i\) and \(A_j\)
describe same concept under different names (synonyms).

Another common redundancy issue in ontologies is redundancy in
hierarchy. This includes sub-class, sub-property and instance
redundancy. Redundancy in hierarchy occurs when ontologist specifies
classes, properties or instances that have hierarchy relations
(\texttt{rdfs:subClassOf}, \texttt{rdfs:subPropertyOf} and
\texttt{owl:instanceOf}) directly or indirectly.

\subsubsection{Consistency}\label{consistency}

In consistency checking of developed ontology the emphasis is on finding
circulatory errors in TBox component of ontology. Circulatory error
occurs when a class is defined as a sub-class or super-class of itself
at any level of hierarchy in the ontology. They can occur with distance
\(0\), \(1\) or \(n\), depending upon the number of relations involved
when traversing the concept down the hierarchy of concepts until we get
the same from where we started traversal. The same also applies for
properties. To evaluate the quality of ontology regarding circulatory
errors the ontology is viewed as graph \(G = (V, E)\), where \(V\) is
set of classes and \(E\) set of \texttt{rdfs:subClassOf} relations.

\subsubsection{Anomaly}\label{anomaly}

Design anomalies prohibit simplicity and maintainability of taxonomic
structures within ontology. They don't cause inaccurate reasoning about
concepts, but point to problematic and badly designed areas in ontology.
Identification and removal of these anomalies should be necessary for
improving the usability and providing better maintainability of
ontology. As OC calculation is concerned there are several components
considered:

\begin{itemize}
\tightlist
\item
  \emph{chain of inheritance} in TBox component,
\item
  \emph{property clumps} and
\item
  \emph{lazy entities} (classes and properties).
\end{itemize}

The notion of chain of inheritance is considered in class hierarchy,
where developer can classify classes as \texttt{rdfs:subClassOf} other
classes up to any level. When such hierarchy of inheritance is long
enough and all classes have no appropriate descriptions in the hierarchy
except inherited child, then ontology suffers from chain of inheritance.
The algorithm for finding and eliminating chains of inheritance is
depicted in Definition \ref{def:chain-of-inheritance}.

\BeginKnitrBlock{definition}[Find chain of inheritance]
\protect\hypertarget{def:chain-of-inheritance}{}{\label{def:chain-of-inheritance}
\iffalse (Find chain of inheritance) \fi{} }

\begin{align*}
&\textbf{findChainOfInheritance} \\
&\quad price = 1 \\
&\quad axiom(C) = [ type, entity, value ] \leftarrow \text{ Axiom of class C} \\
&\quad A(C) = \forall axiom(C) : entity = C \leftarrow \text{ Set of asserted axioms of class C} \\
&\quad A_{\subseteq}^{-} \leftarrow \text{ Set of asserted axioms of class } C \text{ without rdfs:subClassOf axiom} \\
&\quad chainOfInheritance = \{ \} \\
&\quad \text{while } \exists C_i, C_j \in TBox \wedge \exists C_1, C_2, \ldots, C_n \in TBox : (C_j \subseteq C_n \subseteq C_{n-1} \subseteq \ldots \subseteq C_2 \subseteq C_1 \subseteq C_i) \wedge \\
&\qquad ( \forall C_1, C_2, \ldots, C_n : \left | superClass(C_n) \right | = 1 \wedge A_{\subseteq}^{-} (C_n) = \{ \} ) \wedge \left | A_{\subseteq}^{-} (C_i) \right | > 0 \wedge \left | A_{\subseteq}^{-} (C_j) \right | > 0 \text{ then} \\
&\qquad \quad price = price - \frac{n}{n_{\subseteq}^{direct}} \\
&\qquad \quad chainOfInheritance = chainOfInheritance \cup \{ C_i, C_j, \{ C_1, C_2, \ldots, C_n \} \} \\
&\quad \text{end while} \\
&\quad \boldsymbol{chainsOfInheritance} \text{ and } \boldsymbol{price} \\
&\textbf{end}
\end{align*}
\EndKnitrBlock{definition}

The next aspect in design anomalies is property clumps. This problem
occurs when ontologists badly design ontology by using repeated groups
of properties in different class definitions. These groups should be
replaced by an abstract concept composing those properties in all class
definitions where this clump is used. To identify property clumps the
following approach depicted in Definition \ref{def:property-clumps} is
used.

\BeginKnitrBlock{definition}[Find property clumps]
\protect\hypertarget{def:property-clumps}{}{\label{def:property-clumps}
\iffalse (Find property clumps) \fi{} }

\begin{align*}
&\textbf{findPropertyClumps} \\
&\quad price \leftarrow 1 \\
&\quad n_R \leftarrow \text{ Number of properties (datatype and object)} \\
&\quad V \leftarrow \text{ Classes and properties} \\
&\quad E \leftarrow \text{ Links between classes and properties} \\
&\quad propertyClumps = \{ \} \\
&\quad \text{while exist complete bipartite sub-graph } K_{m,n}^{'} \text{ of graph } G(V,E) \\
&\qquad \text{select } K_{m,n}^{''} \text{ from } K_{m,n}^{'} \text{, where } \max (\frac{m^{''} \cdot n^{''}}{m^{''} + n^{''}}) \\
&\qquad propertyClumps = propertyClumps \cup K_{m,n}^{''} \\
&\qquad \text{remove all edges from } G(V, E) \text{ that appear in } K_{m,n}^{''} \\
&\qquad price = price - \frac{ m^{''} \cdot n^{''} - (m^{''} + n^{''}) }{n_R} \\
&\quad \text{end while} \\
&\quad \text{return } \boldsymbol{propertyClumps} \text{ and } \boldsymbol{price} \\
&\textbf{end}
\end{align*}
\EndKnitrBlock{definition}

The last aspect of design anomalies is lazy entities, which is a leaf
class or property in the taxonomy that never appears in the application
and does not have any instances. Eliminating this problem is quite
straightforward; it just requires checking all leaf entities and
verifying whether it contains any instances. In case of existence those
entities should be removed or generalized or instances should be
inserted.

\section{Evaluation}\label{evaluation}

\subsection{Method}\label{method}

\begin{figure}

{\centering \includegraphics[width=0.8\linewidth]{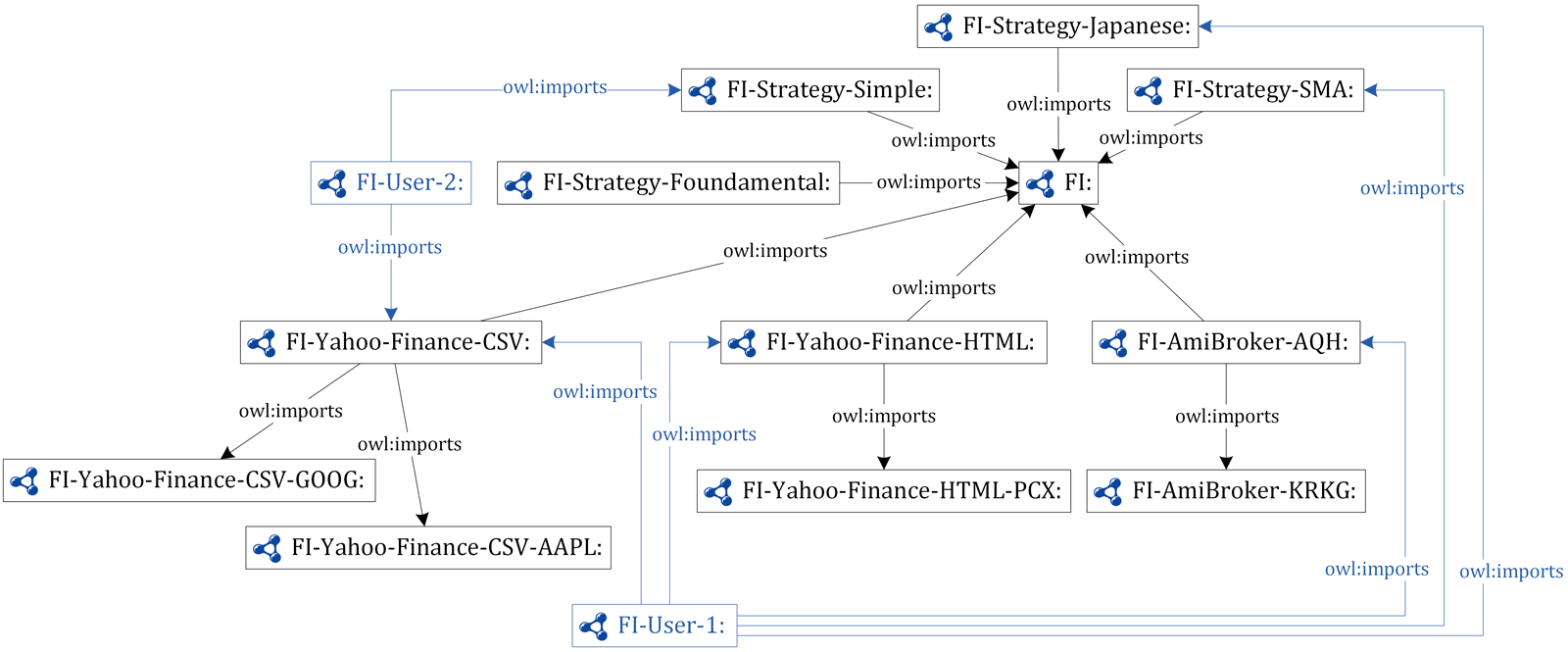}

}

\caption{Financial instruments and trading strategies (FITS)}\label{fig:FITS-ontology}
\end{figure}

The ROD process was evaluated on Financial Instruments and Trading
Strategies (FITS) ontology that is depicted in Figure
\ref{fig:FITS-ontology}.

When building aforementioned ontology one of the requirements was to
follow Semantic Web mantra of achieving as high level of reuse as
possible. Therefore the main building blocks of FITS ontology are all
common concepts about financial instruments. Furthermore every source of
data (e.g., quotes from Yahoo! Finance in a form of CSV files and direct
Web access, AmiBroker trading program format etc.) is encapsulated in a
form of ontology and integrated into FITS ontology. Within every source
of data developer can select which financial instrument is he interested
in (e.g., \texttt{GOOG}, \texttt{AAPL}, \texttt{PCX}, \texttt{KRKG}
etc.). The last and the most important component are financial trading
strategies that developers can define. Every strategy was defined in its
own ontology (e.g., \texttt{FI-Strategy-Simple},
\texttt{FI-Strategy-SMA}, \texttt{FI-Strategy-Japanese} etc.). The
requirement was also to enable open integration of strategies, so
developer can select best practices from several developers and add its
own modification.

Two different approaches in constructing ontology and using it in
aforementioned use case were used. The approach of rapid ontology
development (ROD) was compared to ad-hoc approach to ontology
development, which was based on existing methodologies CommonKADS, OTK
and METHONTOLOGY. With ROD approach the proposed method was used with
tools IntelliOnto and Protégé. The entire development process was
monitored by iteration, where ontology completeness price and number of
ontology elements (classes, properties and axioms with rules) were
followed.

At the end the results included developed ontology, a functional
component and information about the development process by iteration.
The final version of ontology was reviewed by a domain expert, who
confirmed adequateness. At implementation level ontology was expected to
contain about \(250\) to \(350\) axioms of schematic part and numerous
instances from various sources.

\subsection{Results and Discussion}\label{results-and-discussion}

\begin{figure}

{\centering \includegraphics[width=1\linewidth]{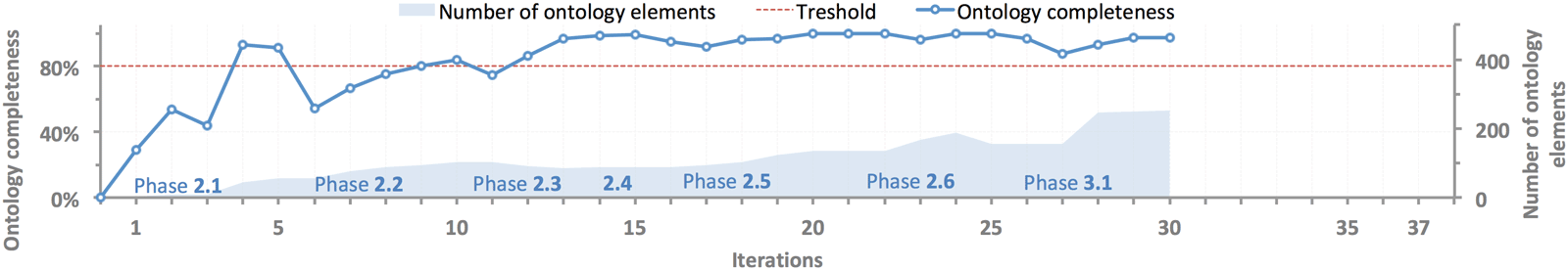}

}

\caption{OC assessment and number of ontology elements through iterations and phases of ROD process}\label{fig:OC-assessment-ROD}
\end{figure}

\begin{figure}

{\centering \includegraphics[width=1\linewidth]{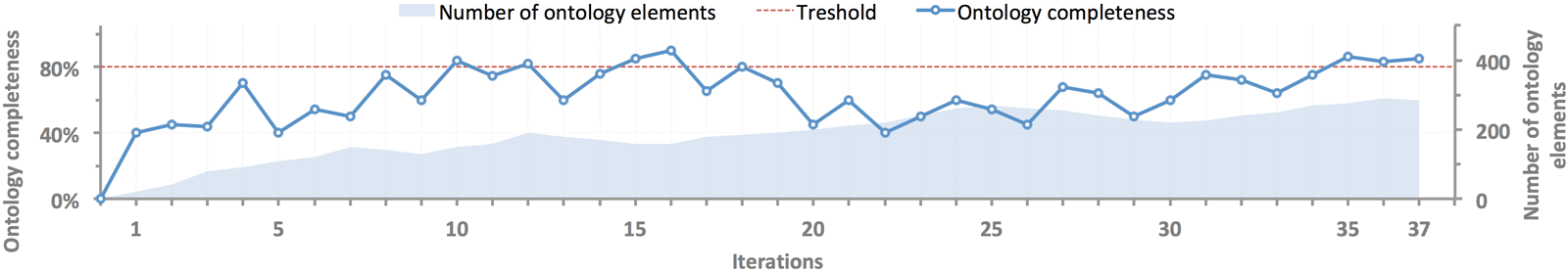}

}

\caption{OC assessment and number of ontology elements through iterations of ad-hoc development process}\label{fig:OC-assessment-ad-hoc}
\end{figure}

The process of ontology creation and exporting it as functional
component was evaluated on FITS ontology and the results are depicted in
Figures \ref{fig:OC-assessment-ROD} and \ref{fig:OC-assessment-ad-hoc}.
Charts represent ontology completeness price and number of ontology
elements regarding to iterations in the process.

Comparing ROD to ad-hoc approach the following conclusions can be drawn:

\begin{itemize}
\tightlist
\item
  the number of iterations to develop required functional component
  using ROD approach \((30)\) is less than using ad-hoc approach
  \((37)\) which results in \(23\%\) less iterations;
\item
  ontology developed with ROD approach is throughout the development
  process more complete and more appropriate for use than in ad-hoc, due
  to continuous evaluation and simultaneous alerts for developers.
\end{itemize}

During the process of ontology construction based on ROD approach the
developer was continuously supported by ontology evaluation and
recommendations for progressing to next steps. When developer entered a
phase and started performing tasks associated with the phase, ontology
completeness was evaluated as depicted in Figure \ref{fig:OC-GUI}. While
OC was less than a threshold value, developer followed instructions for
improving ontology as depicted in Figure \ref{fig:OC-calculation}.
Results of OC evaluation are available in a simple view, where basic
statistics about ontology is displayed (number of concepts, properties,
rules, individuals etc.), progress bar depicting completeness, and
details about evaluation, improvement recommendations and history of
changes. The core element is progress bar that denotes how complete
ontology is and is accompanied with a percentage value. Following are
recommendations for ontology improvement and their gains (e.g., remove
circulatory errors \((+10\%)\), describe concepts in natural language
\((+8\%)\), connect concepts \((+7\%)\) etc.). When improvement is
selected (e.g., remove circulatory errors) the details are displayed
(gain, task and details). The improvement and planned actions are also
clearly graphically depicted on radar chart (see Figure
\ref{fig:OC-GUI}). The shaded area with strong border lines presents
current situation, while red dot shows TO-BE situation if we follow
selected improvement.

When OC price crosses a threshold value (in this experiment \(80\%\)) a
recommendation to progress to a new phase is generated. We can see from
our example that for instance recommendation to progress from phase 2.5
to phase 2.6 was generated in 20th iteration with OC value of
\(91,3\%\), while in 19th iteration OC value was \(76,5\%\).

As Figure \ref{fig:OC-assessment-ROD} depicts ontology completeness
price and number of ontology elements are displayed. While progressing
through steps and phases it's seen that number of ontology elements
constantly grow. On the other hand OC price fluctuate -- it's increasing
till we reach the threshold to progress to next phase and decreases when
entering new phase. Based on recommendations from the system, developer
improves the ontology and OC price increases again. With introduction of
OC steps in ontology development are constantly measured while enabling
developers to focus on content and not technical details (e.g.~language
syntax, best modeling approach etc.).

\section{Conclusions and Future work}\label{conclusion-and-future-work}

Current methodologies and approaches for ontology development require
very experienced users and developers, while we propose ROD approach
that is more suitable for less technically oriented users. With constant
evaluation of developed ontology that is introduced in this approach,
developers get a tool for construction of ontologies with several
advantages:

\begin{itemize}
\tightlist
\item
  the required technical knowledge for ontology modeling is decreased,
\item
  the process of ontology modeling doesn't end with the last successful
  iteration, but continues with post-development activities of using
  ontology as a functional component in several scenarios and
\item
  continuous evaluation of developing ontology and recommendations for
  improvement.
\end{itemize}

In ontology evaluation several components are considered: description,
partition, redundancy, consistency and anomaly. Description of
ontology's components is very important aspect mainly in early stages of
ontology development and includes existence of entities, natural
language descriptions and formal descriptions. This data is furthermore
used for advanced axiom construction in latter stages. Partition errors
deal with omitting important axioms and can be in a form of common
classes, external instances, hierarchy of entities etc. Redundancy deals
with multiple information being inferred more than once and includes
identical formal definition and redundancy in hierarchy. With
consistency the emphasis is on finding circulatory errors, while
anomalies do not cause inaccurate reasoning about concepts, but point to
badly designed areas in ontology. This includes checking for chain of
inheritance, property clumps, lazy entities etc. It has been
demonstrated on a case study from financial trading domain that a
developer can build Semantic Web application for financial trading based
on ontologies that consumes data from various sources and enable
interoperability. The solution can easily be packed into a functional
component and used in various systems.

The future work includes improvement of ontology completeness indicator
by including more semantic checks and providing wider support for
functional components and creating a plug-in for most widely used
ontology editors for constant ontology evaluation. One of the planned
improvements is also integration with popular social networks to enable
developers rapid ontology development, based on reuse.

\end{document}